# Visual Cues in Estimation of Part-To-Whole Comparisons

Stephen Redmond*

National College of Ireland

Accenture plc

**ABSTRACT**

Pie charts were first published in 1801 by William Playfair and have caused some controversy since. Despite the suggestions of many experts against their use, several empirical studies have shown that pie charts are at least as good as alternatives. From Brinton to Few on one side and Eells to Kosara on the other, there appears to have been a hundred-year war waged on the humble pie. In this paper a set of experiments are reported that compare the performance of pie charts and horizontal bar charts with various visual cues. Amazon's Mechanical Turk service was employed to perform the tasks of estimating segments in various part-to-whole charts. The results lead to recommendations for data visualization professionals in developing dashboards.

## 1 INTRODUCTION

In many empirical studies over the years, pie charts will either perform better or just as well as stacked bar charts in part-to-whole comparisons. This is even though a significant number of published authors on the data visualization topic warn against the use of pie charts as being less effective than, say, a stacked bar chart for part-to-whole comparisons.

The pie chart was first published by Playfair [1] in 1801. Since that time, there has been no small amount of discussion and controversy from statisticians, data professionals, designers and journalists. From 1915 [2] through to the modern day, there has been a wide variety of opinions shared in the literature.

A wide range of authors [3] [4] [5] profess the studied opinion that pie charts should not be used as they are not effective. Yet empirical studies [6] [7] [8] appear to show that pie charts are equally effective as the suggested alternatives. It has been suggested [9] [7] that pie charts are better because there is a perceptual anchor used as part of estimation and that pies have naturally occurring angles at 0°, 90°, and 180°.

This long-running continued difference of opinions leads to a question worth exploring and experimenting on: what are the impacts of visual cues on estimation in part-to-whole comparisons.

In this paper an experiment is reported that compares the performance of pie charts and horizontal bar charts with various visual cues as anchors. The Amazon Mechanical Turk service was employed to perform the tasks of estimating segments in these part-to-whole charts.

The results show that pie charts show a better accuracy of estimation than the default style stacked-bar charts based on mean absolute error, even when considering statistical confidence intervals. Additional visual cues do lead to an improvement in accuracy. There is no reason found to reject the hypothesis that a chart's naturally occurring visual anchors assist the accuracy of estimation of segment size. The best visual cue with the bar charts is the traditional quantitative scale, although this is not ideal in all circumstances. These results lead to better recommendations for data visualization professionals in developing dashboards in that they have no reason to reject the use of a pie chart for part-to-whole comparisons.

## 2 RELATED WORK

The Joint Committee of Standard for Graphic Presentation [2] delivers advice to avoid pie charts ("the circle with sectors") as not being desirable. Their advice is that horizontal bars would be more desirable. Brinton, chair of that committee, did not adjust his position on this topic when he later states [3] that a bar chart could be used instead of a sector chart in practically every instance.

Eells [6] performed empirical testing on the subject by having a group of students judge segment sizes of circles and segment lengths of bars, both used as part-to-whole comparison. The results of the study indicate that the pie charts could be read as quickly and more accurately than the bar charts. The study concludes that using circle diagrams to show component parts is "worthy of encouragement". Another element established by the study, and long accepted in the field, is that subjects estimate pie charts by the angle rather than the arc length.

Tufte [4] warns against the use of pie charts and suggests a table is almost always better than using a pie chart. He is especially critical of multiple pie charts used for comparison.

Simkin and Hastie [9] ran studies that appear to confirm that people use perceptual anchors as part of the estimation process. Their study conclusion is that when subjects are estimating part-to-whole segment sizes the visual anchors afforded by position and angle are superior to simple length, as length does not afford any notable visual anchors other than start and end. They found that estimation with pie charts is a special case of the position and angle afforded anchors in that the anchors are at the angles of 0°, 90°, and 180°.

Spence and Lewandowsky [7] follow up on this perceptual anchor point and concur that it is easy to compare the size of a component in the pie chart because of the "imaginary quarters, or halves". Again, their empirical study shows pie charts to be as good as, or superior, to stacked bar charts.

Spence [10] later describes the natural anchors afforded by the pie chart at 0%, 25%, 50%, 75%, 100%, whereas the stacked bar has two, or perhaps three anchors at 0%, 50% and 100%.

---

\* stephen.redmond@ncirl.ie

Few [5] is particularly against the use of pie charts and regularly writes about their use and misuse. He is critical of Spence and Lewandowsky's approach. He does note the advantage that a bar chart has for accurate estimation when there is a quantitative scale.

As well as pie charts, critics will also target variants such as the doughnut chart. Recent work by Skau and Kosara [8] shows that these charts are as effective as a pie. They also show that the pie angle is not the key element for estimation. Kosara has also recently presented experiments [11] [12] comparing various part-to-whole representations, and the pie chart continually performs as well or better than other forms.

Another paper by Kosara [13] calls into question many of the tenets of the field which may be based on aesthetics or the original authors' judgment. He calls on researchers to question and test these assumptions.

The experiments described in this paper seek to empirically test some of the generally held assumptions about part-to-whole charts and sets out to test them.

## 3 EXPERIMENT DESIGN

Four distinct but similar hypotheses are tested in this paper:
1. a bar chart performs better than a pie chart in every instance
2. bar charts with additional visual cues will perform better than without added cues
3. pie charts with additional visual cues will perform better than those without added cues
4. the bar chart with a scale performs better than a bar without a scale

In a similar fashion to previous studies of visual perception [11] [12] [14], the experiments have used crowdsourcing to ask participants to judge a highlighted segment in a chart. No particular cohort of users were selected for and the only assumption is that the worker had enough computer literacy to use the Mechanical Turk system.

To gather the data to test the hypotheses, five values have been chosen to test across all the chart types: 8%, 22%, 28%, 33% and 44%. These represent a good range of segment sizes, from small to almost halfway, and are like the values used by Eells [6] in the 1926 experiments. Six different chart types were used across the experiments. Examples of the chart types can be seen in Figure 1.

For each chart type and value, fifty impressions were shown to workers, giving a total of 1,500 impressions. One worker could not view the same chart and value combination twice.

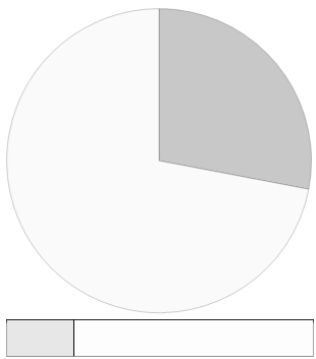

Figure 1: Examples of charts shown to participants. The pie chart segment is 28% and the bar segment is 22%. Participants were asked to judge the size, between 1 and 100, of the darker segment.

316 unique workers participated across all the experiments. The most impressions that any one worker saw was 25.

For each individual task in all the experiments, a worker was shown an impression of one chart that had two segments, a dark segment for the target value, and a lighter segment. the participants were asked the following question:

*"The chart has two segments that both add up to 100%. Please estimate the value of the darker segment. Please enter a whole integer between 1 and 100 only."*

For all experiments, outlier values were removed following the *Tukey Fences* method [15]: based on the interquartile range an outlier to be any observation outside the range $Q_1-k(Q_3-Q_1)$ to $Q_3+k(Q_3-Q_1)$, with $k$=1.5. This left 1,415 valid responses across all the tasks as shown in Table 1.

Table 1. Each presentation was viewed by up to 50 users. Outliers have been removed leaving 1,415 valid responses

|  | 8 | 22 | 28 | 33 | 44 |
|---|---|---|---|---|---|
| Baseline Pie | 40 | 45 | 46 | 47 | 49 |
| Baseline Bar | 49 | 42 | 50 | 49 | 50 |
| Bar with Quartile | 50 | 46 | 48 | 50 | 49 |
| Bar with Decile | 47 | 58 | 45 | 49 | 49 |
| Bar with Scale | 49 | 49 | 49 | 49 | 49 |
| Pie with Quartile | 44 | 44 | 46 | 43 | 45 |

Eells [6] used the mean of absolute variance as the measure to compare performance however, in their seminal paper experimentally evaluating different visual elements, Cleveland and McGill [16] used the log of the absolute error, calculated as $log_2(|judged\ value - true\ value| + c)$. This calculation has also been used in such works as those by Heer and Bostock [14] as well as Kosara and Skau [17]. However, more recently, Kosara [11] [12] has returned to use absolute error for comparison. In this paper, mean of absolute error will be used.

As well as calculating the mean of the absolute error, a statistical 95% confidence level value was calculated to visually show difference following Cumming's New Statistics [18].

## 4 RESULTS

The data was collected and used to test each of the hypotheses. The results of each test are shown below.

### 4.1 Results for hypothesis 1

The first hypothesis test partially replicates that of Eells [6] whereby participants were shown either a bar chart or a pie chart to test the hypothesis that a bar chart performs better than a pie chart in every instance.

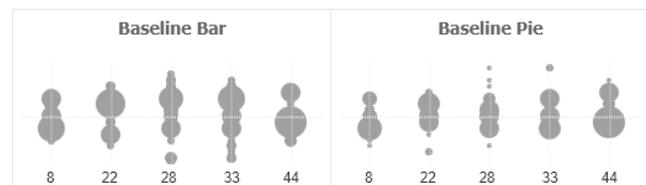

Figure 2: Variation of error values for bar charts and pie charts. The variation of error on the bar charts is greater than for the pies.

The violin plot in Figure 2 indicates a variance of response versus the expected values. Visually, it appears it is wider for the bar charts. Calculating the mean of the absolute error gives the results shown in Table 2.

Table 2. Mean Absolute Error of the baseline set of pie charts versus the baseline set of bar charts.

| Chart Type | Mean Absolute Error |
| --- | --- |
| Baseline Bar | 2.3458 |
| Baseline Pie | 1.7665 |

These results indicate, that the pie chart performed better than the bar chart, and this replicates what Eells reports. Plotting the values, as shown in Figure 3, shows there is a difference between the two, even considering the 95% confidence interval.

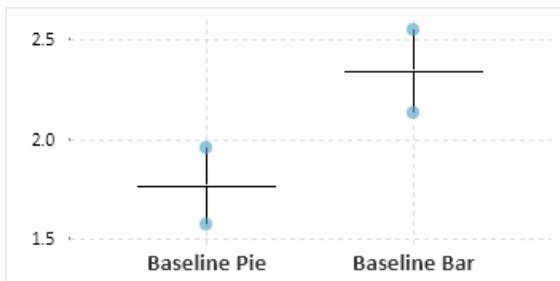

Figure 3: Mean of the absolute error with confidence interval whiskers for baseline bar chart and baseline pie chart. As per Cummings [18], the lack of overlap indicates a significant difference.

Given the replication of the Eells results, the hypothesis of bar superiority can be rejected.

### 4.2 Results for hypothesis 2

To test the hypothesis that bar charts with additional visual cues will perform better than without added cues, bar charts were displayed to participants with additional visual cues. An example of two of these charts with different cues is shown in Figure 4.

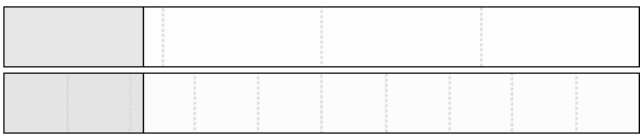

Figure 4: Examples of the bar charts with additional visual cues that were shown to participants. The bar segments are both 22%. The top bar has light colored lines at the quartiles. The lower bar has light colored lines at the deciles. The hypothesis is that these additional cues will improve the accuracy of estimation.

Plotting the mean absolute error and the confidence intervals, as shown in Figure 5, shows the bar with deciles has a significantly lower error than the other charts.

The hypothesis cannot be rejected, but we can see that the number and type of visual cue is important.

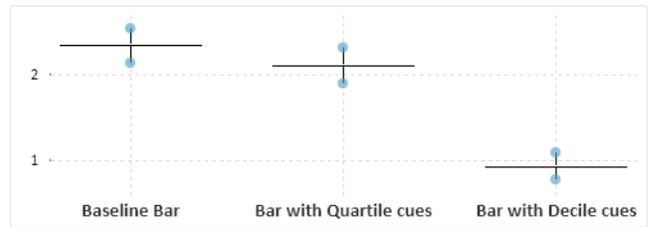

Figure 5: Mean of the absolute error with confidence interval whiskers for baseline bar chart versus bars with quartile visual cues and bars with decile visual cues. The overlap of the baseline and quartile bars indicates no significant difference. There is a significant difference with the decile cues.

### 4.3 Results for hypothesis 3

To test the hypothesis that pie charts with additional visual cues perform better than those without added cues, pie charts were presented to the participants with additional visual cues added at 0%, 25%, 50% and 75%. An example of such as chart, with ticks as visual cues, is show in Figure 6.

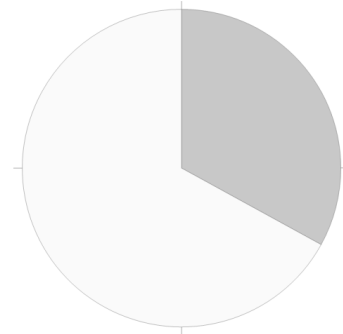

Figure 6: Example of one of the pie charts shown to participants. The pie chart segment is 33%. The additional visual cues are ticks perpendicular to the circle, positioned at 0%, 25%, 50% and 75%

The comparison of the results between the Baseline Pie and the Pie with the Quartile Cues are platted in Figure 7.

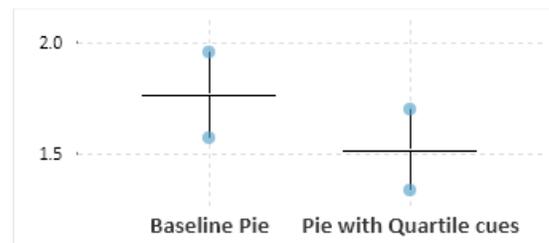

Figure 7: Mean of the absolute error with confidence interval whiskers for baseline pie chart and pie with quartile cue. The overlap of the confidence interval range does not allow a difference to be assumed.

While there is a difference in the mean absolute error, the overlap in the confidence intervals indicates that the different is not significant and the hypothesis can be rejected. This would seem to support the hypothesis that pie charts have natural visual cues, but further experimentation is required.

## 4.4 Results for hypothesis 4

To test the hypothesis that the bar chart with a scale performs better than a bar without a scale, as suggested by Few [5], bar charts with segments were displayed to the participants with an external quantitative scale. An example of a bar chart with a scale is shown in Figure 8.

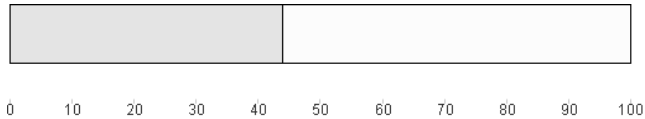

Figure 8: Example of a bar chart with segment at 44% and an external quantitative scale.

The mean of the absolute error for the bar with the scale, as shown in Figure 9, is much lower than the baseline bar, and also lower than the bar with decile visual cues.

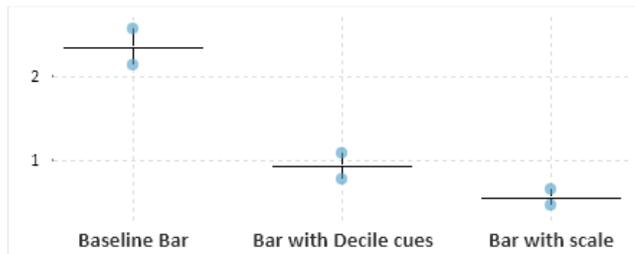

Figure 9: Mean of the absolute error with confidence interval whiskers for baseline bar chart and baseline pie chart. The bar with scale has a lower mean and narrower confidence intervals than even the bar with decile visual cues.

The significance of the difference in mean absolute error that the hypothesis cannot be rejected. The result indicates that a scale does afford better accuracy, although the decile visual cue is also a good option.

## 5 CONCLUSION

This paper set out to explore the impacts of visual anchors on estimation in part-to-whole comparisons, while at the same time empirically testing some generally held assumptions about part-to-whole chart types. Several experiments were carried out to compare the performance of part-to-whole visualizations, specifically pie charts and horizontal bar charts, with various segment sizes, with and without additional visual cues.

The results have shown that the mean absolute error reduces significantly for bar charts as additional visual cues are added. As shown in Figure 10, the variation of the error is also seen to reduce.

The hypothesis that pie charts have no merit and that the bar chart is a better option in almost all cases cannot be substantiated by these experiments. In fact, the opposite is shown to be the case.

Saying that, the assertion of Few that the bar chart is more accurate when it has a quantitative scale can be substantiated from these results. However, it is not always appropriate to provide a scale in certain visualisations, and in those circumstances the bar can be augmented with internal visual cues to act as anchors.

The results show that the baseline charts perform similarly to charts with additional visual cues at the quartiles. This may be evidence that, indeed, there are natural visual anchors in the baseline charts that participants are using. Additional research may be useful here.

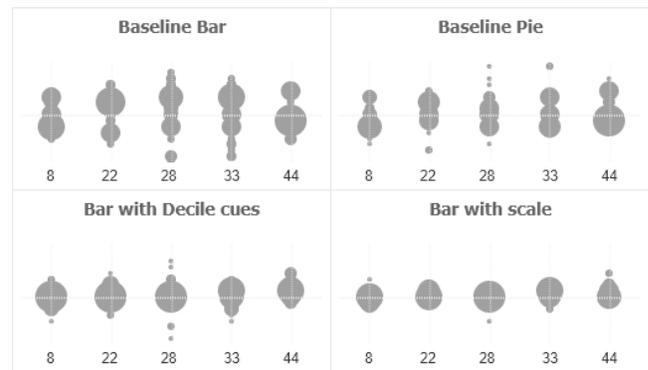

Figure 10: Response variation versus expected value for baseline bar versus bars with scale.

The research presented in this paper supports Kosara's assertion that commonly held truths in data visualisation need to be further researched and tested.

Any data visualisation professional who decides to use a pie chart for a part-to-whole comparison should not feel that they are making an error in that decision.

### ACKNOWLEDGEMENTS

The author would like to thank the reviewers for their comments and contribution to an improved paper.